\documentclass[a4paper]{article}

\usepackage{rotating}
\usepackage{amsmath,amssymb,amsthm}
\usepackage{dsfont}
\usepackage{hyperref}
\usepackage{enumerate}

\usepackage{listings}
\usepackage{xcolor}
\theoremstyle{definition}
\newtheorem{definition}{Definition}

\newcommand{\subexp}[1]{\mathsf{subExp}(#1)}

\newcommand{\findinvariants}{\mathsf{find\_invariants}}
\newcommand{\buildweakenings}{\mathsf{build\_weakenings}}
\newcommand{\coupledweakenings}{\mathsf{coupled\_weakenings}}
\newcommand{\uncoupledweakenings}{\mathsf{uncoupled\_weakenings}}
\newcommand{\loops}{\mathsf{loops}}
\newcommand{\postconditions}{\mathsf{postconditions}}
\newcommand{\outerloops}{\mathsf{outer\_loops}}
\newcommand{\isinvariant}{\mathsf{is\_invariant}}
\newcommand{\aging}{\mathsf{aging}}
\newcommand{\ginpink}{\mathsf{gin\text{-}pink}}

\newcommand{\replace}[1]{\mathsf{replace}(#1)}
\newcommand{\targets}[1]{\mathsf{targets}(#1)}
\newcommand{\variables}[1]{\mathsf{variables}(#1)}

\title{\textsf{Inferring Loop Invariants using Postconditions}}

\author{\textsf{Carlo A. Furia and Bertrand Meyer}}

\date{} 

\begin{document}

\maketitle

\begin{abstract}
One of the obstacles in automatic program proving is to obtain suitable \emph{loop invariants}.
The invariant of a loop is a weakened form of its postcondition (the loop's goal, also known as its contract); the present work takes advantage of this observation by using the postcondition as the basis for invariant inference, using various heuristics such as ``uncoupling'' which prove useful in many important algorithms.
Thanks to these heuristics, the technique is able to infer invariants for a large variety of loop examples.
We present the theory behind the technique, its implementation (freely available for download and currently relying on Microsoft Research's Boogie tool), and the results obtained. 
\end{abstract}

\section{Overview}  \label{sec:overview}
Many of the important contributions to the advancement of program proving have been, rather than grand new concepts, specific developments and simplifications; they have  removed one obstacle after another preventing the large-scale application of  proof techniques to realistic programs built by ordinary programmers in ordinary projects. The work described here seeks to achieve such a practical advance by automatically generating an essential ingredient of proof techniques: \emph{loop invariants}. The key idea is that invariant generation should use not just the text of a loop but its \emph{postcondition}. Using this insight, the $\ginpink$ tool presented here is able to infer loop invariants for non-trivial algorithms including array partitioning (for Quicksort), sequential search, coincidence count, and many others. The tool is available for free download.\footnote{\url{http://se.inf.ethz.ch/people/furia/}}

\subsection{Taking advantage of postconditions}
In the standard Floyd-Hoare approach to program proving, loop invariants are arguably the biggest practical obstacle to full automation of the proof process. Given a routine's specification (contract), in particular its postcondition, the proof process consists of deriving intermediate properties, or \emph{verification conditions}, at every point in the program text. Straightforward techniques yield verification conditions for basic instructions, such as assignment, and basic control structures, such as sequence and conditional. The main difficulty is the loop control structure, where the needed verification condition is a \emph{loop invariant}, which unlike the other cases cannot be computed through simple rules; finding the appropriate loop invariant usually requires human invention.

Experience shows, however, that many programmers find it hard to come up with invariants. This raises the question of devising automatic techniques to infer invariants from the loop's text, in effect extending to loops the mechanisms that successfully compute verification conditions for other constructs. Loops, however, are intrinsically more difficult constructs than (for example) assignments and conditionals, so that in the current state of the art we can only hope for \emph{heuristics} applicable to specific cases, rather than general algorithms guaranteed to yield a correct result in all cases.

While there has been considerable research on loop invariant generation and many interesting results (reviewed in the literature survey of Section~\ref{sec:discussion}), most existing approaches are constrained by a fundamental limitation: to obtain the invariant they only consider the \emph{implementation} of a loop. In addition to raising epistemological problems explained next, such techniques can only try to discover relationships between successive loop iterations; this prevents them from discovering many important classes of invariants.

The distinctive feature of the present work is that it uses the postcondition of a loop for inferring its invariant.  The postcondition is a higher-level view of the loop, describing its goal, and hence allows inferring the correct invariant in many more cases. As will be explained in Section~\ref{sec:invariants-def}, this result follows from the observation that a loop invariant is always a weakened form of the loop's postcondition. Invariant inference as achieved in the present work then relies on implementing a number of \emph{heuristics for weakening postconditions into invariants}; Section~\ref{sec:examples} presents four such heuristics, such as \emph{uncoupling} and \emph{constant relaxation}, which turn out to cover many practical cases.

\subsection{Inferring assertions: the Assertion Inference Paradox}

Any program-proving technique that attempts to infer specification elements (such as loop invariants) from program texts faces a serious epistemological objection, which we may call the Assertion Inference Paradox. 

The Assertion Inference Paradox is a risk of vicious circle.
The goal of program proving is to establish program correctness.
A program is correct if its implementation satisfies its specification; for example a square root routine implements a certain algorithm, intended to reach a final state satisfying the specification that the square of the result  is, within numerical tolerance, equal to the input.
To talk about correctness requires having both elements, the implementation and the specification, and assessing one against the other.
But if we infer the specification from the implementation, does the exercise not become vacuous?
Surely, the proof will succeed, but it will not teach us anything since it loses the fundamental property of independence between the mathematical property to be achieved and the software artifact that attempts to achieve it --- the problem and the solution.

To mitigate the Assertion Inference Paradox objection, one may invoke the following arguments:
\begin{itemize}
\item The Paradox only arises if the goal is to prove correctness.
  Specification inference can have other applications, such as reverse-engineering legacy software.
\item Another possible goal of inferring a specification may be to present it to a programmer, who will examine it for consistency with an intuitive understanding of its intended behavior.
\item Specification inference may produce an inconsistent specification, revealing a flaw in the implementation.
\end{itemize}

For applications to program proving, however, the contradiction remains; an inferred specification not exhibiting any inconsistencies cannot provide a sound basis for a proof process.

For that reason, the present work refrains from attempting specification inference for the principal units of a software system: routines (functions, methods) and those at an even higher level of  granularity (such as classes).
It assumes that these routine specifications are available.
Most likely they will have been written explicitly by humans, although their origin does not matter for the rest of the discussion.

What does matter is that once we have routine specifications, it becomes desirable to infer the specifications of all lower-level constructs (elementary instructions and control structures such as conditionals and loops) automatically.
At those lower levels, the methodological objection expressed by the Assertion Inference Paradox vanishes: the specifications are only useful to express the semantics of implementation constructs, not to guess the software's intent.
The task then becomes: given a routine specification --- typically, a precondition and postcondition --- derive the proof automatically by inferring verification conditions for the constructs used in the routine and proving that the constructs satisfy these conditions.
No vicious circle is created.

For basic constructs such as assignments and conditional instructions, the machinery of Floyd-Hoare logic makes this task straightforward.
The principal remaining difficulty is for loops, since the approach requires exhibiting a \emph{loop invariant}, also known as an \emph{inductive assertion}, and proving that the loop's initialization establishes the invariant and that every execution of the body (when the exit condition is not satisfied) preserves it.

A loop invariant captures the essence of the loop.
Methodologically, it is desirable that programmers devise the invariant while or before devising the loop.
As noted, however, many programmers have difficulty coming up with loop invariants.
This makes invariants an attractive target for automatic inference.

In the present work, then, postconditions are known and loop invariants inferred. The approach has two complementary benefits:
\begin{itemize}
\item It does not raise the risk of circular reasoning since the specification of every program unit is given from the outside, not inferred.
\item Having this specification of a loop's context available gives a considerable boost to loop invariant inference techniques.
  While there is a considerable literature on invariant inference, it is surprising that none of the references with which we are familiar use postconditions.
  Taking advantage of  postconditions makes it possible --- as described in the rest of this paper --- to derive the invariants of many important and sometimes sophisticated loop algorithms that had so far eluded other techniques.
\end{itemize}

\section{Illustrative examples} \label{sec:examples}
This section presents the fundamental ideas behind the loop-invariant generation technique detailed in Section~\ref{sec:algorithm} and demonstrates them on a few examples.
It uses an Eiffel-like \cite{OOSC-2} pseudocode, which facilitates the presentation thanks to the native syntax for contracts and loop invariants.

As already previewed, the core idea is to generate candidate invariants by weakening postconditions according to a few commonly recurring patterns.
The patterns capture some basic ways in which loop iterations modify the program state towards achieving the postcondition.
Drawing both from classic literature \cite{Gri81,BM80} and our own more recent investigations we consider the following fundamental patterns.
\begin{description}
\item[Constant relaxation \cite{BM80,Gri81}:] replace one or more constants by variables.
\item[Uncoupling \cite{BM80}:] replace two occurrences of the same variable each by a different variable.
\item[Term dropping \cite{Gri81}:]  remove a term, usually a conjunct.
\item[Variable aging:] replace a variable by an expression that represents the value the variable had at previous iterations of the loop.
\end{description}
These patterns are then usually used in combination, yielding a number of weakened postconditions.
Each of these candidate invariants is then tested for initiation and consecution (see Section~\ref{sec:invariants-def}) over any loop, and all verified invariants are retained.

The following examples show each of these patterns in action.
The tool described in Sections \ref{sec:algorithm} and \ref{implementation} can correctly infer invariants of these (and more complex) examples.

\subsection{Constant relaxation} \label{sec:constant-relaxation}
Consider the following routine to compute the maximum value in an array.
\begin{lstlisting}[numbers=left,language=OOSC2Eiffel]
  max (A: ARRAY [T]; n: INTEGER): T
     require A.length = n >= 1
     local i: INTEGER
     do
        from i := 0;  Result := A[1];
        until i >= n
        loop
           i := i + 1
           if Result <= A[i] then Result := A[i] end
        end
     ensure forall j::1 <= j &&  j <= n  ==>  A[j] <= Result
\end{lstlisting}

Lines 5--10 may modify variables \lstinline[language=OOSC2Eiffel]|i| and \lstinline[language=OOSC2Eiffel]|Result| but they do not affect input argument \lstinline[language=OOSC2Eiffel]|n|, which is therefore a constant with respect to the loop body.
The \emph{constant relaxation} technique replaces every occurrence of the constant \lstinline[language=OOSC2Eiffel]|n| by a variable \lstinline[language=OOSC2Eiffel]|i|.
The weakened postcondition, \lstinline[language=OOSC2Eiffel]|forall j::1 <= j &&  j <= i ==> A[j] <= Result|, is indeed an invariant of the loop: after every iteration the value of \lstinline[language=OOSC2Eiffel]|Result| is the maximum value of array \lstinline[language=OOSC2Eiffel]|A| over range $[1..i]$.

\subsection{Variable aging} \label{sec:variable-aging}
Sometimes substituting a constant by a variable in the postcondition does not yield any loop invariant because of how the loop body updates the variable.
It may happen that the loop body does not ``use'' the latest value of the substituted variable until the next iteration.
Consider for example another implementation of computing the maximum of an array, which increments variable \lstinline[language=OOSC2Eiffel]|i| \emph{after} using it, so that only the range $[1..i-1]$ of array \lstinline[language=OOSC2Eiffel]|A| has been inspected after every iteration.
\begin{lstlisting}[numbers=left,language=OOSC2Eiffel]
  max_v2 (A: ARRAY [T], n: INTEGER): T
     require A.length = n >= 1
     local i: INTEGER
     do
        from i := 1;  Result := A[1];
        until i > n
        loop
           if Result <= A[i] then Result := A[i] end
           i := i + 1
        end
     ensure  forall j::1 <= j &&  j <= n  ==>  A[j] <= Result
\end{lstlisting}
The \emph{variable aging} heuristics handles these cases by introducing an expression that represents the value of the variable at the previous iteration in terms of its current value.
In the case of routine \lstinline[language=OOSC2Eiffel]|max_v2| it is straightforward that such an expression for variable \lstinline[language=OOSC2Eiffel]|i| is \lstinline[language=OOSC2Eiffel]|i-1.|
The postcondition can be weakened by first replacing variable \lstinline[language=OOSC2Eiffel]|n| by variable \lstinline[language=OOSC2Eiffel]|i| and then by ``aging'' variable \lstinline[language=OOSC2Eiffel]|i| into \lstinline[language=OOSC2Eiffel]|i-1.|
The resulting formula \lstinline[language=OOSC2Eiffel]|forall j::1 <= j &&  j <= i-1 ==> A[j] <= Result| correctly captures the semantics of the loop.

Computing the symbolic value of a variable at the ``previous'' iteration can be quite complex in the general case.
In practice, however, a simple (e.g., flow-insensitive) approximation is often enough to get significant results.
The experiments of Section~\ref{implementation} provide a partial evidence to support this conjecture.

\subsection{Uncoupling} \label{sec:uncoupling}
Consider the task (used as part of the Quicksort algorithm) of partitioning an array \lstinline[language=OOSC2Eiffel]|A| of length \lstinline[language=OOSC2Eiffel]|n| into two parts such that every element of the first part is less than or equal to a given \emph{pivot} value and every element of the second part is greater than or equal to it.
The following contracted routine specifies and implements such task.
\begin{lstlisting}[numbers=left,language=OOSC2Eiffel]
partition (A: ARRAY [T]; n: INTEGER; pivot: T): INTEGER
   require A.length = n >= 1
   local low_index, high_index: INTEGER
   do
     from low_index := 1; high_index := n
     until low_index = high_index
     loop
        from -- no loop initialization
        until low_index = high_index || A[low_index] > pivot
        loop low_index := low_index + 1 end
        from -- no loop initialization
        until low_index = high_index || pivot> A[high_index]
        loop high_index := high_index - 1 end
        A.swap (A, low_index, high_index)
     end
     if pivot <= A[low_index] then
        low_index := low_index - 1
        high_index := low_index
     end
     Result := low_index
   ensure  ( forall k :: 1 <= k  && k < Result + 1 ==> A[k] <= pivot )
	           && ( forall k :: Result < k && k <= n ==> A[k] >= pivot )
\end{lstlisting}

The postcondition consists of the conjunction of two formulas (lines 21 and 22).
If we try to weaken it by replacing constant \lstinline[language=OOSC2Eiffel]|Result| by variable \lstinline[language=OOSC2Eiffel]|low_index| or by variable \lstinline[language=OOSC2Eiffel]|high_index| we obtain no valid loop invariant.
This is because the two clauses of the postcondition should refer, respectively, to portion \lstinline[language=OOSC2Eiffel]|[1..low_index-1]| and \lstinline[language=OOSC2Eiffel]|[high_index+1..n]| of the array.
We achieve this by first uncoupling \lstinline[language=OOSC2Eiffel]|Result|, which means replacing its first occurrence (in line 21) by variable \lstinline[language=OOSC2Eiffel]|low_index| and its second occurrence (in line 22) by variable \lstinline[language=OOSC2Eiffel]|high_index|.
After ``aging'' variable \lstinline[language=OOSC2Eiffel]|low_index| we get the formula:
\begin{lstlisting}
            ( forall k :: 1 <= k && k < low_index ==> A[k] <= pivot )
             &&  ( forall k :: high_index < k && k <= n ==> A[k] >= pivot )
\end{lstlisting}
The reader can check that this indeed a loop invariant of all loops in routine \lstinline[language=OOSC2Eiffel]|partition| and that it allows a straightforward partial correctness proof of the implementation.

\subsection{Term dropping} \label{sec:term-dropping}
The last weakening pattern that we consider consists simply of removing a part of the postcondition.
The formula to be weakened is usually assumed to be in conjunctive normal form, that is, expressed as the conjunction of a few clauses: then term dropping amounts to removing one or more conjuncts.
Going back to the example of \lstinline[language=OOSC2Eiffel]|partition|$\!$, let us drop the first conjunct in the postcondition.
The resulting formula
\begin{lstlisting}[language=OOSC2Eiffel,numbers=none]
               forall k :: Result <  k &&  k <= n  ==>  A[k] >= pivot
\end{lstlisting}
can be further transformed through constant relaxation, so that we end up with a conjunct of the invariant previously obtained by uncoupling: \lstinline[language=OOSC2Eiffel]|forall k::high_index < k && k <= n ==> A[k]>= pivot.|
This conjunct is also by itself an invariant.
In this example term dropping achieved by different means the same result as uncoupling.

\section{Foundations} \label{sec:preliminaries}
Having seen typical examples we now look at the technical choices that support the invariant inference tools.
To decouple the loop-invariant generation technique as much as possible from the specifics of any one programming language, we adopt Boogie from Microsoft Research \cite{Lei08-Boogie2} as our concrete programming language; Section~\ref{sec:boogie} is then devoted to a concise introduction to the features of Boogie that are essential for the remainder.
Sections \ref{sec:invariants-def} and \ref{sec:other-def} introduce definitions of basic concepts and some notational conventions that will be used.
We assume the reader is familiar with standard formal definitions of the axiomatic semantics of imperative programs. 

\subsection{Invariants} \label{sec:invariants-def}
Proving a procedure correct amounts to verifying that:
\begin{enumerate}
\item Every computation terminates.
\item Every call of another procedure is issued only when the preconditions of the callee hold.
\item The postconditions hold upon termination.
\end{enumerate}

It is impossible to establish these facts automatically for all programs but the most trivial ones without additional information provided by the user in the form of annotations.
The most crucial aspect is the characterization of loops, where the expressive power of universal computation lies.
A standard technique to abstract the semantics of any number of iterations of a loop is by means of \emph{loop invariants}.
\begin{definition}[Inductive loop invariant] \label{def:loop-invariant}
Formula $\phi$ is an inductive invariant of loop
\begin{lstlisting}[language=OOSC2Eiffel,numbers=none]
                      from Init until Exit loop Body  end
\end{lstlisting}
iff:
\begin{itemize}
\item \emph{Initiation}: $\phi$ holds after the execution of \lstinline|Init|
\item \emph{Consecution}: the truth of $\phi$ is preserved by every execution of \lstinline|Body| where \lstinline|Exit| does not hold
\end{itemize}
\end{definition}

In the rest of the discussion, inductive invariants will be called just invariants for short.
Note, however, that an invariant in the weaker sense of a property that stays true throughout the loop's execution is not necessarily an \emph{inductive} invariant: in
\begin{lstlisting}[language=OOSC2Eiffel,numbers=none]
                from x := 1 until False loop  x := - x  end
\end{lstlisting}
formula $x \geq -1$ will remain true throughout, but is not considered an inductive invariant because $\{x \geq -1\}\ x := -x\ \{ x \geq -1 \}$ is not a correct Hoare triple.
In the remainder we will deal solely with inductive loop invariants, as is customary in the program proving literature.

From a design methodology perspective, the invariant expresses a weakened form of the postcondition.
More precisely \cite{TouchOfClass,Gri81}, the invariant is a form of the postcondition that applies to a subset of the data, and satisfies the following three properties:
\begin{enumerate}
\item It is \emph{strong enough} to yield the postcondition when combined with the exit condition (which states that the loop has covered the \emph{entire} data).

\item It is \emph{weak enough} to make it easy to write an algorithm (the loop initialization \lstinline|Init|$\!$) that will satisfy the invariant on a subset (usually empty or trivial) of the data.

\item It is \emph{weak enough} to make it easy to write an algorithm (the  loop body \lstinline|Body|) that, given that the invariant holds on a subset of the data that is not the entire data, extends it to cover a slightly larger subset.
\end{enumerate}

``Easy'', in the last two conditions, means ``much easier than solving the entire original problem''.
The loop consists of an approximation strategy that starts with the initialization, establishing the invariant, then retains the invariant while extending the scope by successive approximations to an ever larger set of the input through repeated executions of the loop body, until it hits the exit condition, signaling that it now covers the entire data and hence satisfies the postcondition.
This explains that the various strategies of Section~\ref{sec:examples}, such as constant relaxation and uncoupling, are heuristics for weakening the postcondition.

\subsection{Boogie} \label{sec:boogie}
Boogie, now in its second version, is both an intermediate verification language and a verification tool. 

The Boogie language combines a typed logical specification language with an in-the-small imperative programming language with variables, procedures, contracts, and annotations.
The type system comprises a few basic primitive types as well as type constructors such as one- and two-dimensional arrays.
It supports a relatively straightforward encoding of object-oriented language constructs.
Indeed, Boogie is part of the Spec\# programming environment; mappings have been defined for other programming languages, including Eiffel \cite{Tsc09-Julian} and C \cite{SXSP07}.
This suggests that the results described here can be generalized to many other contexts.

The Boogie tool verifies conformance of a procedure to its specification by generating verification conditions (VC) and feeding them to an automated theorem prover (the standard one being Z3).
The outcome of a verification attempt can be successful or unsuccessful.
In the latter case the tool provides some feedback on what might be wrong in the procedure, in particular by pointing out what contracts or annotations it could not verify.
Verification with Boogie is sound but incomplete: a verified procedure is always guaranteed to be correct, while an unsuccessful verification attempt might simply be due to limitations of the technology.

\subsubsection{The Boogie specification language} \label{sec:boogie-spec-lang}
The Boogie specification language is essentially a typed predicate calculus with equality and arithmetic.
Correspondingly, formulas --- that is, logic expressions --- are built by combining atomic constants, logic variables, and program variables with relational and arithmetic operators, as well as with Boolean connectives and quantifiers.
For example, the following formula (from Section~\ref{sec:constant-relaxation}) states that no element in array $X$ within positions $1$ and $n$ is larger than $v$: in other words, the array has maximum $v$.
\begin{lstlisting}
                forall j : int :: 1 <= j &&  j <= n  ==>  X[j] <= v
\end{lstlisting}
The syntactic classes $Id$, $Number$, and $Map$ represent constant and variable identifiers, numbers, and mappings, respectively.

Complex formulas and expressions can be postulated in \emph{axioms} and parameterized by means of logic \emph{functions}.
Functions are a means of introducing encapsulation and genericity for formulas and complex expressions.
For example, the previous formula can be parameterized into function \lstinline|is_max| with the following signature and definition:
\begin{lstlisting}
      function is_max (m: int, A: array int, low: int, high: int)
         returns ( bool )
         { forall j : int :: low <= j &&  j <= high  ==>  A[j] <= m }
\end{lstlisting}

Axioms constrain global constants, variables, and functions; they are useful to supply Boogie with domain knowledge to facilitate inference and guide the automated reasoning over non-trivial programs.
In certain situations it might for example be helpful to introduce the property that if an array has maximum $m$ over range $[low..high]$ and the element in position $high + 1$ is smaller than $m$ then $m$ is also the maximum over range $[low..high+1]$.
The following Boogie axiom will express this:
\begin{lstlisting}
         axiom ( forall m: int, A: array int, low: int, high: int ::
              is_max(m, A, low, high) &&  A[high + 1] < m
                     ==>  is_max (m, A, low, high+1) )
\end{lstlisting}

\subsubsection{The Boogie programming language} \label{sec:boogie-prog-lang}
A Boogie program is a collection of \emph{procedures}.
Each procedure consists of a signature, a \emph{specification} and (optionally) an \emph{implementation} or body.
The signature gives the procedure a name and declares its formal input and output arguments.
The specification is a collection of contract clauses of three types: \emph{frame conditions}, \emph{preconditions}, and \emph{postconditions}.

A frame condition, introduced by keyword \lstinline|modifies|, consists of a list of global variables that can be modified by the procedure; it is useful in evaluating the side-effects of procedure call within any context.
A precondition, introduced by the keyword \lstinline|requires|, is a formula that is required to hold upon procedure invocation.
A postcondition, introduced by the keyword \lstinline|ensures|, is a formula that is guaranteed to hold upon successful termination of the procedure.
For example, procedure \lstinline|max_v2|{}, computing the maximum value in an array $A$ given its size $n$, has the following specification.
\begin{lstlisting}
procedure max_v2 (A: array int, n: int) returns (m: int)
   requires n >=  1;
   ensures is_max (m, A, 1, n);
\end{lstlisting}

The implementation of a procedure consists of a declaration of local variables, followed by a sequence of (possibly labeled) program statements.
Figure \ref{fig:body-syntax} shows a simplified syntax for Boogie statements.
Statements of class \emph{Annotation} introduce checks at any program point: an \emph{assertion} is a formula that must hold of every execution that reaches it for the program to be correct and an \emph{assumption} is a formula whose validity at the program point is postulated.
Statements of class \emph{Modification} affect the value of program variables, by nondeterministically drawing a value for them (\lstinline|havoc|), assigning them the value of an expression (\lstinline|:=|), or calling a procedure with actual arguments (\lstinline|call|).
The usual conditional \lstinline|if| statement controls the execution flow.
Finally, the \lstinline|while| statement supports loop iteration, where any loop can be optionally annotated with a number of \emph{Invariants} (see Section~\ref{sec:invariants-def}).
Boogie can check whether Definition \ref{def:loop-invariant} holds for any user-provided loop invariant.

\begin{figure}[!tb]
\begin{scriptsize}
\begin{center}
\begin{tabular}{rcl}
Statement    &  ::=  &  Assertion $\mid$ Modification \\
             & $\mid$&  ConditionalBranch $\mid$ Loop  \\
Annotation   &  ::=  &  \textbf{assert} Formula $\mid$ \textbf{assume} Formula  \\
Modification &  ::=  &  \textbf{havoc} VariableId $\mid$ VariableId := Expression \\
             &   $|$   &  \textbf{call} [ VariableId$^+$ := ] ProcedureId ( Expression* ) \\
ConditionalBranch &   ::=   &  \textbf{if} ( Formula ) Statement* [ \textbf{else} Statement* ] \\
Loop       &   ::=   &  \textbf{while} ( Formula ) Invariant* Statement*  \\
Invariant  &   ::=   &  \textbf{invariant} Formula
\end{tabular}
\end{center}
\end{scriptsize}
\caption{Simplified abstract syntax of Boogie statements}
\label{fig:body-syntax}
\end{figure}

The implementation of procedure \lstinline|max_v2| is:
\begin{lstlisting}
   var i: int;
   i := 1;  m := A[1];
   while (i <=  n)
   {
	   if (m <= A[i])  { m := A[i]; }
		 i := i + 1;
   }
\end{lstlisting}

While the full Boogie language includes more types of statement, any Boogie statement can be desugared into one of those in Figure \ref{fig:body-syntax}.
In particular, the only looping construct we consider is the structured \lstinline|while|; this choice simplifies the presentation of our loop invariant inference technique and makes it closer as if it was defined directly on a mainstream high-level programming language.
Also, there is a direct correspondence between Boogie's \lstinline|while| loop and Eiffel's \lstinline[language=OOSC2Eiffel]|from ... until| loop, used in the examples of Section~\ref{sec:examples} and the definitions in Section~\ref{sec:invariants-def}.

\subsection{Notational conventions} \label{sec:other-def}
$\subexp{\phi, SubType}$ denotes the set of sub-ex\-pres\-sions of formula $\phi$ that are of syntactic type $SubType$.
For example, $\subexp{\text{\lstinline!is_max(v,X,1,n)!}, Map}$ denotes all mapping sub-ex\-pres\-sions in \lstinline!is_max(v,X,1,n)!, that is only \lstinline!X[j]!.

$\replace{\phi, old, new, *}$ denotes the formula obtained from $\phi$ by replacing every occurrence of sub-ex\-pres\-sion $old$ by expression $new$.
Similarly, $\replace{\phi, old,\linebreak new, n}$ denotes the formula obtained from $\phi$ by replacing only the $n$-th occurrence of sub-ex\-pres\-sion $old$ by expression $new$, where the total ordering of sub-ex\-pres\-sions is given by a pre-order traversal of the expression parse tree.
For example, $\replace{\text{\lstinline!is_max(v,X,1,n)!}, j, h, *}$ is:
\begin{lstlisting}
               forall h : int :: low <= h &&  h <= high  ==>  A[h] <= m
\end{lstlisting}
while $\replace{\text{\lstinline!is_max(v,X,1,n)!}, j, h, 4}$ is:
\begin{lstlisting}
               forall j : int :: low <= j &&  j <= high  ==>  A[h] <= m
\end{lstlisting}

Given a while loop $\ell$: \lstinline|while ( ... ) { Body }|, $\targets{\ell}$ denotes the set of its \emph{targets}: variables (including mappings) that can be modified by its \emph{Body}; this includes global variables that appear in the \lstinline|modifies| clause of called procedures.

Given a \lstinline|procedure foo|, $\variables{\textit{foo}}$ denotes the set of all variables that are visible within \lstinline|foo|, that is its locals and any global variable.

A loop $\ell'$ is \emph{nested} within another loop $\ell$, and we write $\ell' \prec \ell$, iff $\ell'$ belongs to the \emph{Body} of $\ell$.
Notice that if $\ell' \prec \ell$ then $\targets{\ell'} \subseteq \targets{\ell}$.
Given a \lstinline|procedure foo|, its \emph{outer} while loops are those in its body that are not nested within any other loop.

\section{Generating loop invariants from postconditions} \label{sec:algorithm}
This section presents the loop invariant generation algorithm in some detail. 


\begin{figure}
\begin{lstlisting}[language=EiffelLikePseudocode]
$\findinvariants$ ( a_procedure: PROCEDURE )
               : SET_OF [FORMULA]
  do
    Result := $\emptyset$
    for each  post in $\postconditions$(a_procedure) do
      for each loop in $\outerloops$(a_procedure) do
        -- compute all weakenings of post
        -- according to chosen strategies
        weakenings := $\buildweakenings$(post, loop)
        for each  formula in weakenings do
          for each any_loop in $\loops$(a_procedure) do
            if $\isinvariant$(formula, any_loop) then
              Result := Result $\cup\:\{ formula \}$
\end{lstlisting}
\caption{Procedure $\findinvariants$}
\label{fig:find-invariants}
\end{figure}

\subsection{Main algorithm}
The pseudocode in Figure \ref{fig:find-invariants} describes the main algorithm for loop-invariant generation.
The algorithm operates on a given procedure and returns a set of formulas that are invariant of \emph{some} loop in the procedure.
Every postcondition $post$ among all postconditions $\postconditions(a\_procedure)$ of the procedure is considered separately (line 5).
This is a coarse-grained yet effective way of implementing the \emph{term-dropping} strategy outlined in Section~\ref{sec:term-dropping}: the syntax of the specification language supports splitting postconditions into a number of conjuncts, each introduced by the \lstinline|ensures| keyword, hence each of these conjuncts is weakened in isolation.
It is reasonable to assume that the splitting into \lstinline|ensures| clauses performed by the user separates logically separated portions of the postcondition, hence it makes sense to analyze each of them separately.
This assumption might fail, of course, and in such cases the algorithm can be enhanced to consider more complex combinations of portions of the postcondition.
However, one should only move to this more complex analysis if the basic strategy --- which is often effective --- fails.
This enhancement belongs to future work.

The algorithm of Figure \ref{fig:find-invariants} then considers every \emph{outer} while loop (line 6).
For each of them, it computes a set of weakenings of postcondition $post$ (line 9) according to the heuristics of Section~\ref{sec:examples}.
It then examines each weakening to determine if it is invariant to \emph{any} loop in the procedure under analysis (lines 10--13), and finally it returns the set \lstinline[language=EiffelLikePseudocode]|Result| of weakened postconditions that are invariants to some loop.
For $\isinvariant(formula, loop)$, the check consists of verifying whether initiation and consecution hold for $formula$ with respect to $loop$, according to Definition \ref{def:loop-invariant}.
How to do this in practice is non-trivial, because loop invariants of different loops within the same procedure may interact in a circular way: the validity of one of them can be established only if the validity of the others is known already and \emph{vice versa}.

This was the case of \lstinline|partition| presented in Section~\ref{sec:uncoupling}.
Establishing consecution for the weakened postcondition in the outer loop requires knowing that the same weakened postcondition is invariant to each of the two internal while loops (lines 8--13) because they belong to the body of the outer while loop.
At the same time, establishing initiation for the first internal loop requires that consecution holds for the outer while loop, as every new iteration of the external loop initializes the first internal loop.
Section~\ref{implementation} discusses a straightforward, yet effective, solution to this problem.


\begin{figure}
\begin{lstlisting}[language=EiffelLikePseudocode]
$\buildweakenings$ ( post: FORMULA; loop: LOOP )
                 : SET_OF [FORMULA]
  do
   Result := $\{post\}$
   all_subexpressions := $\subexp{post, Id} \;\cup\;$
                         $\subexp{post, Number} \;\cup\;$
                         $\subexp{post, Map}$
   for each constant in all_subexpressions$ \setminus \targets{loop}$ do
      for each variable in $\targets{loop}$ do
         Result := Result $\;\cup\;$
            $\coupledweakenings$(post, constant, variable) $\;\cup\;$
            $\uncoupledweakenings$(post, constant, variable)
\end{lstlisting}
\caption{Procedure $\buildweakenings$}
\label{fig:build-weakenings}
\end{figure}

\subsection{Building weakened postconditions}
Algorithm $\buildweakenings(post, loop)$, described in Figure \ref{fig:build-weakenings}, computes a set of weakened versions of postcondition formula $post$ with respect to outer while loop $loop$.
It first includes the unchanged postcondition among the weakenings (line 4).
Then, it computes (lines 5--7) a list of sub-ex\-pres\-sions of $post$ made of atomic variable identifiers (syntactic class $Id$), numeric constants (syntactic class $Number$) and references to elements in arrays (syntactic class $Map$).
Each of these sub-ex\-pres\-sions that $loop$ does not modify (i.e., it is not one of its targets) is a $constant$ with respect to the loop.
The algorithm then applies the \emph{constant relaxation} heuristics of Section~\ref{sec:constant-relaxation} by relaxing $constant$ into any $variable$ among the $loop$'s targets (lines 8--9).
More precisely, it computes two sets of weakenings for each pair $\langle constant, variable \rangle$: in one \emph{uncoupling}, described in Section~\ref{sec:uncoupling}, is also applied (lines 12 and 11, respectively).

The fact that any target of the loop is a candidate for substitution justifies our choice of considering only outer while loops: if a loop $\ell'$ is nested within another loop $\ell$ then $\targets{\ell'} \subseteq \targets{\ell}$, so considering outer while loops is a conservative approximation that does not overlook any possible substitution.

\begin{figure}
\begin{lstlisting}[language=EiffelLikePseudocode]
$\coupledweakenings$
      ( post: FORMULA; constant, variable: EXPRESSION )
      : SET_OF [FORMULA]
  do
   Result := $\replace{post, constant, variable, *}$
   aged_variable := $\aging$(variable, loop)
   Result := Result $\;\cup\;$
                 $\replace{post, constant, aged\_variable, *}$
\end{lstlisting}
\caption{Procedure $\coupledweakenings$}
\label{fig:coupled-weakenings}
\end{figure}

\subsection{Coupled weakenings}
The algorithm in Figure \ref{fig:coupled-weakenings} applies the constant relaxation heuristics to postcondition $post$ without uncoupling.
Hence, relaxing $constant$ into $variable$ simply amounts to replacing every occurrence of $constant$ by $variable$ in $post$ (line 5); i.e., $\replace{post, constant, variable, *}$ using the notation introduced in Section~\ref{sec:other-def}.
Afterward, the algorithm applies the other \emph{aging} heuristics (introduced in Section~\ref{sec:variable-aging}): it computes the ``previous'' value of $variable$ in an execution of $loop$ (line 6) and it substitutes the resulting expression for $constant$ in $post$ (lines 7--8).

While the implementation of function $\aging$ could be very complex we adopt the following unsophisticated approach.
For every possible acyclic execution path in $loop$, we compute the symbolic value of $variable$ with initial value $v_0$ as a symbolic expression $\epsilon(v_0)$.
Then we obtain $\aging(variable, loop)$ by solving the equation $\epsilon(v_0) = variable$ for $v_0$, for every execution path.\footnote{Note that $\aging(variable, loop)$ is in general a set of expressions, so the notation at lines 6--8 in Figure \ref{fig:coupled-weakenings} is a convenient shorthand.}
For example, if the loop simply increments $variable$ by one, then $\epsilon(v_0) = v_0 + 1$ and therefore $\aging(variable, loop) = variable - 1$.
Again, while the example is unsophisticated it is quite effective in practice; indeed, most of the times it is enough to consider simple increments or decrements of $variable$ to get a ``good enough'' aged expression.

\subsection{Uncoupled weakenings}
The algorithm of Figure \ref{fig:uncoupled-weakenings} is a variation of the algorithm of Figure \ref{fig:coupled-weakenings} applying the \emph{uncoupling} heuristics outlined in Section~\ref{sec:uncoupling}.
It achieves this by considering every occurrence of $constant$ in $post$ separately when performing the substitution of $constant$ into $variable$ (line 6).
Everything else is as in the non-uncoupled case; in particular, aging is applied to every candidate for substitution.

This implementation of uncoupling relaxes one occurrence of a constant at a time.
This is not the most general implementation of uncoupling, as in some cases it might be useful to substitute different occurrences of the same constant by different variables.
This was the case of \lstinline|partition| discussed in Section~\ref{sec:uncoupling}, where relaxing two occurrences of the same constant \lstinline[language=OOSC2Eiffel]|Result| into two different variables was needed in order to get a valid invariant.
Section~\ref{sec:term-dropping} showed, however, that the term-dropping heuristics would have made this ``double'' relaxation unnecessary for the procedure.

\begin{figure}
\begin{lstlisting}[language=EiffelLikePseudocode]
$\uncoupledweakenings$
      (post: FORMULA; constant, variable: EXPRESSION)
      : SET_OF [FORMULA]
  do
   Result := $\emptyset$;  index := 1
   for each occurrence of constant in post do
     Result := Result $\;\cup\;$
        $\{\replace{post, constant, variable, index}\}$
     aged_variable := $\aging$(variable, loop)
     Result := Result $\;\cup\;$
        $\{\replace{post, constant, aged\_variable, index}\}$
     index := index + 1
\end{lstlisting}
\caption{Procedure $\uncoupledweakenings$}
\label{fig:uncoupled-weakenings}
\end{figure}

	

\section{Implementation and experiments} \label{implementation}
We developed a command-line tool code-named $\ginpink$ (Generation of INvariants by PostcondItioN weaKening) implementing in Eiffel the loop-invariant inference technique described in Section~\ref{sec:algorithm}.
While we plan to integrate $\ginpink$ into EVE (the Eiffel Verification Environment\footnote{\url{http://eve.origo.ethz.ch}}) where it will analyze the code resulting from the translation of Eiffel into Boogie \cite{Tsc09-Julian}, its availability as a stand-alone tool makes it possible to use it for languages other than Eiffel provided a Boogie translator is available.

$\ginpink$ applies the algorithm of Figure \ref{fig:find-invariants} to some selected procedure in a Boogie file provided by the user.
After generating all weakened postconditions it invokes the Boogie tool to determine which of them is indeed an invariant: for every candidate invariant $I$, a new copy of the original Boogie file is generated with $I$ declared as invariant of \emph{all} loops in the procedure under analysis.
It then repeats the following until either Boogie has verified all current declarations of $I$ in the file or no more instances of $I$ exist in the procedure:
\begin{enumerate}
  \item Use Boogie to check whether the current instances of $I$ are verified invariants, that is they satisfy initiation and consecution.
  \item If any candidate fails this check, comment it out of the file.
\end{enumerate}
In the end, the file contains all invariants that survive the check, as well as a number of invariants that could not be checked in the form of comments.
If no invariant survives or the verified invariants are unsatisfactory, the user can still manually inspect the generated files to see if verification failed due to the limited reasoning capabilities of Boogie.

When generating candidate invariants, $\ginpink$ does not apply all heuristics at once but it tries them incrementally, according to user-supplied options.
Typically, the user starts out with just constant relaxation and checks if some non-trivial invariant is found.
If not, the analysis is refined by gradually introducing the other heuristics --- and thus increasing the number of candidate invariants as well.
In the experiments below we briefly discuss how often and to what extent this is necessary in practice.

\paragraph{Experiments}
Table \ref{tab:experiments} summarizes the results of a number of experiments with $\ginpink$ with a number of Boogie procedures obtained from Eiffel code.
We carried out the experimental evaluation as follows.
First, we collected examples from various sources \cite{BM80,Gri81,PG02,LM09,BLW08} and we manually completed the annotations of every algorithm with full pre and postconditions as well as with any loop invariant or intermediate assertion needed in the correctness proof.
Then, we coded and tried to verify the annotated programs in Boogie, supplying some background theory to support the reasoning whenever necessary.
The latest Boogie technology cannot verify certain classes of properties without a very sophisticated \emph{ad hoc} background theory or without abstracting away certain aspects of the implementation under verification.
For example, in our implementation of Bubblesort, Boogie had difficulties proving that the output is a permutation of the input.
Correspondingly, we omitted the (few) parts of the specification that Boogie could not prove even with a detailedly annotated program.
Indeed, ``completeness'' (full functional correctness) should not be a primary concern, because its significance depends on properties of the prover (here Boogie), orthogonal to the task of inferring invariants.
Finally, we ran $\ginpink$ on each of the examples after commenting out all annotations except for pre and postconditions (but leaving the simple background theories in); in a few difficult cases (discussed next) we ran additional experiments with some of the annotations left in.
After running the tests, we measured the relevance of every automatically inferred invariant: we call an inferred invariant \emph{relevant} if the correctness proof needs it.
Notice that our choice of omitting postcondition clauses that Boogie cannot prove does not influence relevance, which only measures the fraction of inferred invariants that are useful for proving correctness.

For each experiment, Table \ref{tab:experiments} reports: the name of the procedure under analysis; the length in lines of codes (the whole file including annotations and auxiliary procedures and, in parentheses, just the main procedure); the total number of loops (and the maximum number of nested loops, in parentheses); the total number of variables modified by the loops (scalar variables/array or map variables); the number of weakened postconditions (i.e., candidate invariants) generated by the tool; how many invariants it finds; the number and percentage of verified invariants that are relevant; the total run-time of $\ginpink$ in seconds; the source (if any) of the implementation and the annotations.
The experiments where performed on a PC equipped with an Intel Quad-Core 2.40 GHz CPU and 4 Gb of RAM, running Windows XP as guest operating system on a VirtualBox virtual machine hosted by Ubuntu GNU/Linux 9.04 with kernel 2.6.28.

Most of the experiments already succeeded with the application of the most basic weakening techniques.
Procedure \emph{Coincidence Count} is the only case that required a more sophisticated uncoupling strategy where two occurrences of the same constant within the same formula were weakened to two different aged variables.
This resulted in an explosion of the number of candidate invariants and consequently in an experiment running for over an hour.

A few programs raised another difficulty, due to Boogie's need for user-supplied loop invariants to help automated deduction.
Boogie cannot verify any invariant in \emph{Shortest Path} or \emph{Topological Sort} without additional invariants obtained by means other than the application of the algorithm itself.
On the other hand, the performance with programs \emph{Array Stack Reversal} and \emph{Dutch National Flag} improves considerably if user-supplied loop invariants are included, but fair results can be obtained even without any such annotation.
Table \ref{tab:experiments} reports both experiments, with and without user-supplied annotations.

More generally, Boogie's reasoning abilities are limited by the amount of information provided in the input file in the form of axioms and functions that postulate sound inference rules for the program at hand.
We tried to limit this amount as much as possible by developing the necessary theories before tackling invariant generation.
In other words, the axiomatizations provided are enough for Boogie to prove functional correctness with a properly annotated program, but we did not strengthen them only to ameliorate the inference of invariants.
We believe that a richer axiomatization could have removed the need for user-supplied invariants in the programs considered.

\begin{table}
\begin{scriptsize}
\begin{center}
\begin{tabular}{crcrrrrrc}
\textsc{Procedure}  &  \textsc{LOC}  &  \textsc{\# lp.} &  \textsc{m.v.}  & \textsc{cnd.}  &  \textsc{inv.}  &  \textsc{rel.}  &  \textsc{T.}  & \textsc{Src.} \\
\hline
\emph{Array Partitioning} (v1)           &  58 (22)   &   1 (1)  &  2/1 &   38 &  9 & 	3 ( 33\%) &   93  \\
\emph{Array Partitioning} (v2)           &  68 (40)  & 3(2) & 2/1 &   45 &	 2 &	2 (100\%) &  205 & \cite{BM80} \\
\emph{Array Stack Reversal}              &  147 (34)  & 2 (1) & 1/2 & 134 &	 4 &	2 ( 50\%) &  529 &  \\
\emph{Array Stack Reversal} (ann.)  &  147 (34)  & 2 (1) &  1/2 & 134 &	 6 &	4 ( 67\%) &  516 &  \\
\emph{Bubblesort}                       &  69 (29)   & 2(2) & 2/1 &  14 &	 2 &	2 (100\%) &   65 & \cite{PG02} \\
\emph{Coincidence Count}                 &  59 (29)   & 1 (1) & 3/0 & 1351 &	 1 &	1 (100\%) & 4304 & \cite{LM09} \\
\emph{Dutch National Flag}                        &  77 (43)   & 1 (1) & 3/1 &  42 &	10 &	2 ( 20\%) &  117 & \cite{Why-manual} \\
\emph{Dutch National Flag} (ann.)            &  77 (43)   & 1 (1) & 3/1 &  42 &	12 &	4 ( 33\%) &  122 & \cite{Why-manual} \\
\emph{Majority Count}                    &  48 (37)   & 1 (1) &  3/0 & 23 &	 5 &	2 (40\%) &   62 &  \cite{BM91,Morgan94} \\
\emph{Max of Array} (v1)                 &  27 (17)   & 1 (1) & 2/0 &  13 &	 1 &	1 (100\%) &   30 &  \\
\emph{Max of Array} (v2)                 &  27 (17)   & 1 (1) & 2/0 &   7 &	 1 &	1 (100\%) &   16 &  \\
\emph{Plateau}                           &  53 (29)   & 1 (1) & 3/0 &  31 &	 6 &	3 ( 50\%) &  666 & \cite{Gri81} \\
\emph{Sequential Search} (v1)            &  34 (26)   & 1 (1) & 3/0 &  45 &	 9 &	5 ( 56\%) &  120 &  \\
\emph{Sequential Search} (v2)            &  29 (21)   & 1 (1) & 3/0 &  24 &	 6 &	6 (100\%) &   58 &  \\
\emph{Shortest Path} (ann.)         &  57 (44)   & 1 (1) &   1/4 & 23 &	 2 &	2 (100\%) &   53 & \cite{BLW08} \\
\emph{Stack Search}                      &  196 (49)  & 2 (1) & 1/3 &  102 &	 3 &	3 (100\%) &  300 &  \\
\emph{Sum of Array}                      &  26 (15)   & 1 (1) & 2/0 &  13 &	 1 &	1 (100\%) &   44 &  \\
\emph{Topological Sort}  (ann.)          &  65 (48)   & 1 (1) & 2/4 &  21 &	 3 &	2 (67\%) &   101 &  \cite{TouchOfClass} \\
\emph{Welfare Crook}                     &  53 (21)   & 1 (1) & 3/0 &  20 &	 2 &	2 (100\%) &  586 & \cite{Gri81}
\end{tabular}
\end{center}
\end{scriptsize}
\caption{Experiments with $\ginpink$.}
\label{tab:experiments}
\end{table}

\section{Discussion and related work} \label{sec:discussion}

\subsection{Discussion}
The experiments in Section~\ref{implementation} provide a partial, yet significant, assessment of the practicality and effectiveness of our technique for loop invariant inference.
Two important factors to evaluate any inference technique deserve comment: relevance of the inferred invariants and scalability to larger programs.

A large portion of the invariants retrieved by $\ginpink$ are relevant --- i.e., required for a functional correctness proof --- and complex --- i.e., involving first-order quantification over several program elements.
To some extent, this is unsurprising because deriving invariants from postconditions ensures by construction that they play a central role in the correctness proof and that they are at least as complex as the postcondition.

As for scalability to larger programs, the main problem is the combinatorial explosion of the candidate invariants to be checked as the number of variables that are modified by the loop increases.
In properly engineered code, each routine should not be too large or call too many other routines. 
The empirical observations mentioned in \cite[Sec.~9]{KV09} seem to support this assumption, which ensures that the candidate invariants do not proliferate and hence the inference technique can scale within reasonable limits.
The examples of Section~\ref{implementation} are not trivial in terms of length and complexity of loops and procedures, if the yardstick is well-modularized code.
On the other hand, there is plenty of room for finessing the application order of the various weakening heuristics in order to analyze the most ``promising'' candidates first; the Houdini approach \cite{FL01} might also be useful in this context.
The investigation of these aspects belongs to future work.

\subsection{Limitations}
Relevant invariants obtained by postcondition weakening are most of the times significant, practically useful, and complementary to a large extent to the categories that are better tackled by other methods (see next sub-section).
Still, the postcondition weakening technique cannot obtain \emph{every} relevant invariant.
Failures have two main different origins: conceptual limitations and shortcomings of the currently used technology.

The first category covers invariants that are not expressible as weakening of the postcondition.
This is the case, in particular, whenever an invariant refers to a local variable whose final state is not mentioned in the postcondition.
For example, the postcondition of procedure \lstinline|max| in Section~\ref{sec:constant-relaxation} does not mention variable \lstinline|i| because its final value \lstinline|n| is not relevant for the correctness.
Correspondingly, invariant \lstinline|i <= n| --- which is involved in the partial correctness proof --- cannot be obtained by postcondition weakening.
A potential solution to these conceptual limitations is two-fold: on the one hand, many of these invariants that escape postcondition weakening can be obtained reliably with other inference techniques that do not require postconditions --- this is the case of invariant \lstinline|i <= n| in procedure \lstinline|max| which is retrieved automatically by Boogie.
On the other hand, if we can augment postconditions with complete information about local variables, the weakening approach can have a chance to work.
In the case of \lstinline|max|, a dynamic technique could suggest the supplementary postcondition \lstinline|i <= n && i >= n| which would give the sought invariant by dropping the second conjunct.

Shortcomings of the second category follow from limitations of state-of-the-art automated theorem provers, which prevent reasoning about certain interesting classes of algorithms.
As a simple example, consider the following implementation of Newton's algorithm for the square root of a real number, more precisely the variant known as the Babylonian algorithm (we ignore numerical precision issues) \cite{BM80}:
\begin{lstlisting}[language=OOSC2Eiffel,numbers=none]
     square_root (a: REAL): REAL
        require a >= 0
        local y: REAL
        do
           from Result := 1;  y := a
           until Result = y
           loop
              Result := (Result + y)/2
              y := a / Result
           end
        ensure Result >= 0  &&  Result * Result = a
\end{lstlisting}
Postcondition weakening would correctly find invariant \lstinline[language=OOSC2Eiffel]|Result * y = a| (by term dropping and uncoupling), but Boogie cannot verify that it is an invariant because the embedded theorem prover Z3 does not handle reasoning about properties of products of numeric variables \cite{LM09}.
If we can verify by other means that a candidate is indeed an invariant, the postcondition weakening technique of this paper would be effective over additional classes of programs.

\subsection{Related work}
The amount of research work on the automated inference of invariants is for\-mi\-da\-ble and spread over more than three decades; this reflects the cardinal role that invariants play in the formal analysis and verification of programs.
This section outlines a few fundamental approaches and provides some evidence that this paper's technique is complementary, in terms of kinds of invariants inferred, to previously published approaches.
For more references, in particular regarding software engineering applications, see the ``related work'' section of \cite{ECGN01}.

\paragraph{Static methods}
Historically, the earliest methods for invariant inference where \emph{static} as in the pioneering work of Karr \cite{Kar76}.
Abstract interpretation and the constraint-based approach are the two most widespread frameworks for static invariant inference (see also \cite[Chap.~12]{BM07-book}).

\emph{Abstract interpretation} is, roughly, a symbolic execution of programs over abstract domains that over-approximates the semantics of loop iteration.
Since the seminal work by Cousot and Cousot \cite{CC77}, the technique has been updated and extended to deal with features of modern programming languages such as object-orientation and heap memory-management (e.g., \cite{Log04,CL05}).

\emph{Constraint-based} techniques rely on sophisticated decision procedures over non-trivial mathematical domains (such as polynomials or convex polyhedra) to represent concisely the semantics of loops with respect to certain template properties.

Static methods are sound --- as is the technique introduced in this paper --- and often complete with respect to the class of invariants that they can infer.
Soundness and completeness are achieved by leveraging the decidability of the underlying mathematical domains they represent; this implies that the extension of these techniques to new classes of properties is often limited by undecidability.
In fact, state-of-the-art static techniques can mostly infer invariants in the form of ``well-behaving'' mathematical domains such as linear inequalities \cite{CH78,CSS03}, polynomials \cite{SSM04,RCK07}, restricted properties of arrays \cite{BMS06,BHIKV09,HHKV10}, and linear arithmetic with uninterpreted functions \cite{BHMR07}.
Loop invariants in these forms are extremely useful but rarely sufficient to prove full functional correctness of programs.
In fact, one of the main successes of abstract interpretation has been the development of sound but incomplete tools \cite{BCC+03-Astree} that can verify the absence of simple and common programming errors such as division by zero or void dereferencing.
Static techniques for invariant inference are now routinely part of modern static checkers such as ESC/Java \cite{FLL+02-ESCJava}, Boogie/Spec\# \cite{Lei08-Boogie2}, and Why/Krakatoa/Caduceus \cite{FM07}.

The technique of the present paper is complementary to most static techniques in terms of the kinds of invariant that it can infer, because it derives invariants directly from postconditions.
In this respect ``classic'' static inference and our inference by means of postcondition weakening can fruitfully work together to facilitate functional verification; to some extent this happens already when complementing Boogie's built-in facilities for invariant inference with our own technique.

\cite{PV04-spin,Jan07,dCGG09,LQGVW09,KV09} are the approaches that, for different reasons, share more similarities with ours.
To our knowledge, \cite{PV04-spin,Jan07,dCGG09,LQGVW09} are the only other works applying a static approach to derive loop invariants from annotations.
\cite{Jan07} relies on user-provided assertions nested within loop bodies and essentially tries to check whether they hold as invariants of the loop.
This does not release the burden of writing annotations nested within the code, which is quite complex as opposed to providing only contracts in the form of pre and postconditions.
In practice, the method of \cite{Jan07} works only when the user-provided annotations are very close to the actual invariant; in fact the few examples where the technique works are quite simple and the resulting invariants are usually obtainable by other techniques that do not need annotations.
\cite{dCGG09} briefly discusses deriving the invariant of a \emph{for} loop from its postcondition, within a framework for reasoning about programs written in a specialized programming language.
\cite{LQGVW09} also leverages specifications to derive intermediate assertions, but focusing on lower-level and type-like properties of pointers.
On the other hand, \cite{PV04-spin} derives candidate invariants from postconditions in a very different setting than ours, with symbolic execution and model-checking techniques.

Finally, \cite{KV09} derives complex loop invariants by first encoding the loop semantics as recurring relations and then instructing a rewrite-based theorem prover to try to remove the dependency on the iterator variable(s) in the relations.
It shares with our work a practical attitude that favors powerful heuristics over completeness and leverages state-of-the-art verification tools to boost the inference of additional annotations.

\paragraph{Dynamic methods}
More recently, dynamic techniques have been applied to invariant inference.
The Daikon approach of Ernst et al.~\cite{ECGN01} showed that dynamic inference is practical and sprung much derivative work (e.g., \cite{PE04,CTS08,PCM09} and many others).
In a nutshell, the Daikon approach consists in testing a large number of candidate properties against several program runs; the properties that are not violated in any of the runs are retained as ``likely'' invariants.
This implies that the inference is not sound but only an ``educated guess'': dynamic invariant inference is to static inference what testing is to program proofs.
Nonetheless, just like testing is quite effective and useful in practice, dynamic invariant inference is efficacious and many of the guessed invariants are indeed sound.

Our approach shares with the Daikon approach the idea of guessing a candidate invariant and testing it \emph{a posteriori}.
There is an obvious difference between our approach, which retains only invariants that can be soundly verified, and dynamic inference techniques, which rely on a finite set of tests.
A deeper difference is that Daikon guesses candidate invariants almost blindly, by trying out a pre-defined set of user-provided templates (including comparisons between variables, simple inequalities, and simple list comprehensions).
On the contrary, our technique assumes the availability of contracts (and postconditions in particular) and leverages it to restrict quickly the state-space of search and get to good-quality loop invariants in a short time.
As it is the case for static techniques, dynamic invariant inference methods can also be usefully combined with our technique, in such a way that invariants discovered by dynamic methods boost the application of the postcondition-weakening approach.

\paragraph{Program construction}
Classical formal methods for program construction \cite{Dij76,Gri81,BM80,Morgan94} have first described the idea of deriving loop invariants from postconditions.
Several of the heuristics that we discussed in Section~\ref{sec:examples} are indeed a rigorous and detailed rendition of some ideas informally presented in \cite{BM80,Gri81}.
In addition, the focus of the seminal work on program construction is to derive systematically an implementation from a complete functional specification.
In this paper the goal is instead to enrich the assertions of an already implemented program and to exploit its contracts to annotate the code with useful invariants that facilitate a functional correctness proof.

\section{Conclusion and future work} \label{conclusion}
As we hope to have shown, taking advantage of postconditions makes it possible to obtain loop invariants through effective techniques --- not as predictable as the algorithms that yield verification conditions for basic constructs such as assignments and conditionals, but sufficiently straightforward to be applied by tools, and yielding satisfactory results in many practical cases.

The method appears general enough, covering most cases in which a programmer with a strong background in Hoare logic would be able at some effort to derive the invariant, but a less experienced one would be befuddled. So it does appear to fill what may be the biggest practical obstacle to automatic program proving.

The method requires that the programmer (or a different person, the ``proof engineer'', complementing the programmer's work, as testers traditionally do) provide the postcondition for every routine. As has been discussed in Section~\ref{sec:overview}, we feel that this is a reasonable expectation for serious development, reflected in the Design by Contract methodology. For some people, however, the very idea of asking programmers or other members of a development team to come up with contracts of any kind is unacceptable. With such an a priori assumption, the results of this paper will be of little practical value; the only hope is to rely on invariant inference techniques that require the program only (complemented, in approaches such as Daikon, by test results and a repertoire of invariant patterns).

Some of the results that the present approach yields (sometimes trivially) when it is applied manually, are not yet available through the tools used in the current implementation of $\ginpink$. Although undecidability results indicate that program proving will never succeed in all possible cases, it is fair to expect that many of these limitations --- such as those following from Z3's current inability to handle properties of products of variables --- will go away as proof technology continues to progress.

We believe that the results reported here can play a significant
role in the effort to make program proving painless and even matter-of-course.
So in addition to the obvious extensions --- making sure the method covers all effective patterns of postcondition weakening, and taking advantage of progress in theorem prover technology --- our most important task for the near future is to integrate the results of this article, as unobtrusively as possible for the practicing programmer, in the background of a verification environment for contracted object-oriented software components.


\end{document}